\documentclass[]{spie}  

\usepackage{amsmath,amsfonts,amssymb,siunitx}
\usepackage{graphicx}
\usepackage[colorlinks=true, allcolors=blue]{hyperref}
\usepackage{glossaries}
\usepackage{soul}
\usepackage[colorinlistoftodos]{todonotes}  
\usepackage{soul}
\usepackage[caption=false,font=footnotesize]{subfig}
\usepackage{booktabs,tabularx}
\newcolumntype{Y}{>{\centering\arraybackslash}X}

\title{The Deformable Mirror Demonstration Mission (DeMi) CubeSat: optomechanical design validation and laboratory calibration}

\author[a]{Gregory Allan}
\author[a]{Ewan S. Douglas}
\author[a]{Derek Barnes}
\author[a]{Mark Egan}
\author[b]{Gabor Furesz}
\author[a]{Warren Grunwald}
\author[a,c]{Jennifer Gubner}
\author[a]{Christian Haughwout}
\author[a]{Bobby G. Holden}
\author[a]{Paula do Vale Pereira}
\author[a]{Abigail J. Stein}
\author[a,d]{Kerri L. Cahoy}
\affil[a]{Massachusetts Institute of Technology, Department of Aeronautics and Astronautics, Cambridge, MA, USA}
\affil[b]{Massachusetts Institute of Technology, Kavli Institute for Astrophysics and Space Research, Cambridge, MA, USA}
\affil[c]{Wellesley College, Wellesley, MA, USA}
\affil[d]{Massachusetts Institute of Technology, Department of Earth, Atmospheric, and Planetary Science, Cambridge, MA, USA}

\authorinfo{Further author information: (Send correspondence to G.A. or E.S.D.)\\G.A: E-mail: gregallan@mit.edu\\  E.S.D: E-mail:douglase@bu.edu}

\newacronym{AU}{AU}{Astronomical Unit [1.5e11 m]}  
\newacronym{pc}{pc}{parsec}
\newacronym{mas}{mas}{milliarcsecond}
\newacronym{nm}{nm}{Nanometer}
\newacronym{CTE}{CTE}{coefficient of thermal expansion}

\newacronym{smc}{SMC}{Small Magellanic Cloud}
\newacronym{lmc}{LMC}{Large Magellanic Cloud}
\newacronym{ism}{ISM}{interstellar medium}
\newacronym{mw}{MW}{Milky Way}
\newacronym{epseri}{$\epsilon$ Eri}{Epsilon Eridani}
\newacronym{EKB}{EKB}{Edgeworth-Kuiper Belt}

\newacronym{CFR}{CFR}{Complete Frequency Redistribution}

\newacronym{nasa}{NASA}{National Aeronautics and Space Agency}
\newacronym{esa}{ESA}{European Space Agency}
\newacronym{omi}{OMI}{\textit{Optical Mechanics Inc.}}
\newacronym{gsfc}{GSFC}{\gls{nasa} Goddard Space Flight Center}
\newacronym{stsci}{STScI}{Space Telescope Science Institute}
\newacronym{nsroc}{NSROC}{\gls{nasa} Sounding Rocket Operations Contract}
\newacronym{wff}{WFF}{\gls{nasa} Wallops Flight Facility}
\newacronym{wsmr}{WSMR}{White Sands Missile Range}

\newacronym{irac}{IRAC}{Infrared Array Camera}
\newacronym[plural=CCDs, firstplural=charge-coupled devices (CCDs)]{ccd}{CCD}{charge-coupled device}
\newacronym{DM}{DM}{Deformable Mirror}
\newacronym{MCP}{MCP}{ Microchannel Plate }
\newacronym{ipc}{IPC}{Image Proportional Counter}
\newacronym{cots}{COTS}{Commercial Off-The-Shelf}
\newacronym{ISR}{ISR}{Incoherent Scatter Radar }
\newacronym{atcamera}{AT}{Angle Tracker}
\newacronym{MEMS}{MEMS}{microelectromechanical systems}
\newacronym{QE}{QE}{quantum efficiency}
\newacronym{RTD}{RTD}{Resistance Temperature Detector}
\newacronym{PID}{PID}{Proportional-Integral-Derivative}
\newacronym{PRNU}{PRNU}{photo response non-uniformity}
\newacronym{DSNU}{PRNU}{dark signal non-uniformity}
\newacronym{CMOS}{CMOS}{complementary metal–oxide–semiconductor}
\newacronym{TRL}{TRL}{technology readiness level}

\newacronym{FOV}{FOV}{field-of-view}
\newacronym{NIR}{NIR}{near-infrared}
\newacronym{PV}{PV}{Peak-to-Valley}
\newacronym{MRF}{MRF}{Magnetorheological finishing}
\newacronym{AO}{AO}{Adaptive Optics}
\newacronym{TTP}{TTP}{tip, tilt, and piston}
\newacronym{FPS}{FPS}{fine pointing system}
\newacronym{SHWFS}{SHWFS}{Shack-Hartmann Wavefront Sensor}
\newacronym{OAP}{OAP}{off-axis parabola}
\newacronym{LGS}{LGS}{laser guide star}

\newacronym{acs}{ACS}{Attitude Control System}
\newacronym{ADCS}{ADCS}{Attitude Determination and Control System}
\newacronym{orsa}{ORSA}{Ogive Recovery System Assembly}
\newacronym{gse}{GSE}{Ground Station Equipment}
\newacronym{FSM}{FSM}{Fast Steering Mirror}
\newacronym{SWAP}{SWAP}{size, weight, and power}
\newacronym{COTS}{COTS}{commercial, off-the-shelf}

\newacronym{WFS}{WFS}{wavefront sensor}
\newacronym{LSI}{LSI}{Lateral Shearing Interferometer}
\newacronym{VVC}{VVC}{Vector Vortex Coronagraph}
\newacronym{VNC}{VNC}{Visible Nulling Coronagraph}
\newacronym{CGI}{CGI}{Coronagraph Instrument}
\newacronym{IWA}{IWA}{Inner Working Angle}
\newacronym{OWA}{OWA}{Outer Working Angle}
\newacronym{NPZT}{N-PZT}{Nuller piezoelectric transducer}
\newacronym{OPD}{OPD}{Optical Path Difference}
\newacronym{WFCS}{WFCS}{Wavefront Control System}
\newacronym{ZWFS}{ZWFS}{Zernike wavefront sensor}
\newacronym{SPC}{SPC}{Shaped Pupil Coronagraph}
\newacronym{HLC}{HLC}{Hybrid-Lyot Coronagraph}

\newacronym{HST}{HST}{ Hubble Space Telescope}
\newacronym{GPS}{GPS}{Global Positioning System}
\newacronym{ISS}{ISS}{International Space Station}
\newacronym[description=Advanced CCD Imaging Spectrometer]{acis}{ACIS}{Advanced \gls{ccd} Imaging Spectrometer}
\newacronym{stis}{STIS}{\textit{Space Telescope Imaging Spectrograph}}
\newacronym{mcp}{MCP}{Microchannel Plate}
\newacronym{jwst}{JWST}{$\textit{James Webb Space Telescope}$}
\newacronym{fuse}{FUSE}{$\textit{FUSE}$}
\newacronym{galex}{GALEX}{$\textit{Galaxy Evolution Explorer}$}
\newacronym{spitzer}{Spitzer}{$\textit{Spitzer Space Telescope}$}
\newacronym{mips}{MIPS}{Multiband Imaging Photometer for \gls{spitzer}}
\newacronym{gissmo}{GISSMO}{Gas Ionization Solar Spectral Monitor}
\newacronym{iue}{IUE}{International Ultraviolet Explorer}
\newacronym{spinr}{SPINR}{$\textit{Spectrograph for Photometric Imaging with Numeric Reconstruction}$}
\newacronym{imager}{IMAGER}{$\textit{Interstellar Medium Absorption Gradient Experiment Rocket}$}
\newacronym{TPF-C}{TPF-C}{Terrestrial Planet Finder Coronagraph}
\newacronym{RAIDS}{RAIDS}{Atmospheric and Ionospheric Detection System }
\newacronym{mama}{MAMA}{Multi-Anode Microchannel Array}
\newacronym{ATLAST}{ATLAST}{Advanced Technology Large Aperture Space Telescope}
\newacronym{PICTURE}{PICTURE}{Planet Imaging Concept Testbed Using a Rocket Experiment}
\newacronym{LITES}{LITES}{Limb-imaging Ionospheric and Thermospheric
Extreme-ultraviolet Spectrograph}
\newacronym{LBT}{LBT}{Large Binocular Telescope}
\newacronym{LBTI}{LBTI}{Large Binocular Telescope Interferometer}
\newacronym{KIN}{KIN}{Keck Interferometer Nuller}
\newacronym{SHARPI}{SHARPI}{Solar High-Angular Resolution Photometric Imager}
\newacronym{IRAS}{IRAS}{Infrared Astronomical Satellite}
\newacronym{HARPS}{HARPS}{High Accuracy Radial velocity Planetary}
\newacronym{hstSTIS}{STIS}{Space Telescope Imaging Spectrograph}
\newacronym{spitzerIRAC}{IRAC}{Infrared Array Camera}
\newacronym{spitzerMIPS}{MIPS}{Multiband Imaging Photometer for Spitzer}
\newacronym{spitzerIRS}{IRS}{Infrared Spectrograph}
\newacronym{CHARA}{CHARA}{Center for High Angular Resolution Astronomy}
\newacronym{wfirst-afta}{WFIRST-AFTA}{Wide-Field InfrarRed Survey
Telescope-Astrophysics Focused Telescope Assets}
\newacronym{GPI}{GPI}{Gemini Planet Imager}
\newacronym{WFIRST}{WFIRST}{Wide-Field InfrarRed Survey Telescope}
\newacronym{HabEx}{HabEx}{Habitable Exoplanet Imaging Mission}
\newacronym{LUVOIR}{LUVOIR}{Large UV/Optical/Infrared Surveyor}
\newacronym{FGS}{FGS}{Fine Guidance Sensor}
\newacronym{STIS}{STIS}{Space Telescope Imaging Spectrograph}
\newacronym{MGHPCC}{MGHPCC}{Massachusetts Green High Performance
Computing Center}
\newacronym{WISE}{WISE}{Wide-field Infrared Survey Explorer}
\newacronym{ALMA}{ALMA}{Atacama Large Millimeter Array}

\newacronym{AURIC}{AURIC}{The Atmospheric Ultraviolet Radiance Integrated Code} 
\newacronym{FFT}{FFT}{Fast Fourier Transform  }
\newacronym{MODTRAN}{MODTRAN   }{ MODerate resolution atmospheric TRANsmission }
\newacronym{idl}{IDL}{$\textit {Interactive Data Language}$}
\newacronym[sort=NED,description=NASA/IPAC Extragalactic Database]{ned}{NED}{\gls{nasa}/\gls{ipac} Extragalactic Database}
\newacronym{iraf}{IRAF}{Image Reduction and Analysis Facility}
\newacronym{wcs}{WCS}{World Coordinate System}
\newacronym{pegase}{PEGASE}{$\textit{Projet d'Etude des GAlaxies par Synthese Evolutive}$}
\newacronym{dirty}{DIRTY}{$\textit{DustI Radiative Transfer, Yeah!}$}
\newacronym{CUDA}{CUDA}{Compute Unified Device Architecture}

\newacronym{MSIS}{MSIS}{Mass Spectrometer Incoherent Scatter Radar}
\newacronym{nmf2}{$N_m$}{F2-Region Peak density}
\newacronym{hmf2}{$h_m$}{F2-Region Peak height}
\newacronym{H}{$H$}{F2-Region Scale Height}

\newacronym{isr}{ISR}{Incoherent Scatter Radar}
\newacronym[description=TLA Within Another Acronym]{twaa}{TWAA}{\gls{tla} Within Another Acronym}
\newacronym[plural=SNe, firstplural=Supernovae (SNe)]{sn}{SN}{Supernova}
\newacronym{EUV}{EUV}{Extreme-Ultraviolet }
\newacronym{EUVS}{EUVS}{\gls{EUV} Spectrograph}
\newacronym{F2}{F2}{Ionospheric Chapman F Layer }
\newacronym{F10.7}{F10.7}{ 10.7 cm radio flux [10$^{-22}$ W m$^{-2}$ Hz$^{-1}$]  }
\newacronym{FUV}{FUV}{ Far-Ultraviolet }
\newacronym{IR}{IR}{Infrared}
\newacronym{MUV}{MUV}{Mid-Ultraviolet }
\newacronym{NUV}{NUV}{Near-Ultraviolet }
\newacronym{O$^+$}{O$^+$}{Singly Ionized Oxygen  Atom }
\newacronym{OI}{OI}{Neutral Atomic Oxygen Spectroscopic State }
\newacronym{OII}{OII}{Singly Ionized Atomic Oxygen Spectroscopic State }
\newacronym{PSF}{PSF}{Point Spread Function}
\newacronym{$R_E$}{$R_E$}{ Earth Radii [$\approx$ 6400 km]  }
\newacronym{RV}{RV}{Radial Velocity}
\newacronym{UV}{UV}{Ultraviolet }
\newacronym{WFE}{WFE}{Wavefront Error}
\newacronym{sed}{SED}{Spectral Energy Distribution}
\newacronym{nir}{NIR}{near-infrared}
\newacronym{mir}{MIR}{mid-infrared}
\newacronym{ir}{IR}{infrared}
\newacronym{uv}{UV}{ultraviolet}
\newacronym[plural=PAHs, firstplural=Polycyclic Aromatic Hydrocarbons (PAHs)]{pah}{PAH}{Polycyclic Aromatic Hydrocarbon}
\newacronym{obsid}{OBSID}{Observation Identification}
\newacronym{SZA}{SZA}{Solar Zenith Angle}
\newacronym{TLE}{TLE}{Two Line Element set}
\newacronym{DOF}{DOF}{degrees-of-freedom}
\newacronym{PZT}{PZT}{lead zirconate titanate}

\newacronym{PCA}{PCA}{Principal Component Analysis}
\newacronym{fwhm}{FWHM}{Full-Width-Half Maximum}
\newacronym{RMS}{RMS}{root mean squared}
\newacronym{RMSE}{RMSE}{root mean squared error}
\newacronym{MCMC}{MCMC}{Marcov chain Monte Carlo}
\newacronym{DIT}{DIT}{Discrete Inverse Theory}
\newacronym{SNR}{SNR}{signal-to-noise ratio}
\newacronym{PSD}{PSD}{Power Spectral Density}

 \newacronym{LEO}{LEO}{low Earth orbit}
\begin{document} 
\maketitle

\begin{abstract}
Coronagraphs on future space telescopes will require precise wavefront correction to detect Earth-like exoplanets near their host stars. High-actuator count microelectromechanical system (MEMS) deformable mirrors provide wavefront control with low size, weight, and power. The Deformable Mirror Demonstration Mission (DeMi) payload will demonstrate a 140 actuator MEMS \gls{DM} with \SI{5.5}{\micro\meter} maximum stroke. We present the flight optomechanical design, lab tests of the flight wavefront sensor and wavefront reconstructor, and simulations of closed-loop control of wavefront aberrations. We also present the compact flight \gls{DM} controller, capable of driving up to 192 actuator channels at 0-250V with 14-bit resolution. Two embedded Raspberry Pi 3 compute modules are used for task management and wavefront reconstruction. The spacecraft is a 6U CubeSat (30 cm x 20 cm x 10 cm) and launch is planned for  2019. 
\end{abstract}


\section{INTRODUCTION}
\label{sec:intro}  
\subsection{Background and Motivation}
 Deformable Mirrors (DMs) with actuator counts as high as 128 across the pupil are an essential technology for high-contrast imaging using future large space observatories \cite{pueyo_luvoir_2017}. \gls{MEMS} \gls{DM}s are particularly well suited to the task because of their high actuator density. Their lower \gls{SWAP} when compared to piezo-electric, electrostrictive, or voice coil designs also may allow  easier accommodation in spacecraft, and their low actuator mass makes them resilient to launch-induced vibrations. \gls{DM}s are also useful for other in-space applications, including optical communication and wide-field scanning telescopes \cite{scott_wide_2010} as well as deployable \cite{champagne_cubesat_2014}, self-assembling \cite{underwood_using_2015}, and other types of reconfigurable optical systems. 

\gls{MEMS} optical devices have flown in space previously, including a single micro-mirror on the \gls{MEMS} Telescope for Extreme Lightning (MTEL) \cite{nam_telescope_2008,jeon_performance_2016}, and a microshutter array \cite{fleming_calibration_2013}. A high-actuator count \gls{MEMS} \gls{DM} has been operated briefly during a sub-orbital sounding rocket flight \cite{douglas_wavefront_2018}, but additional validation is needed to show that \gls{MEMS} \gls{DM}s are suitable for long-term in-space use on an operational telescope. The devices will need to withstand radiation effects, spacecraft charging, long-term temperature cycling, and extended operation in vacuum. Long-term environmental effects are difficult to replicate in a lab environment, so an actual \gls{LEO} mission is needed to validate the technology. The preparation for this mission will also provide risk reduction for the challenging development of flight software, electronics hardware, and mechanical packaging necessary to operate a DM in space.

The DeMi mission\cite{cahoy_wavefront_2013, douglas_design_2017-1,marinan_improving_2016} is a 6U CubeSat  which will raise the \gls{TRL} of \gls{MEMS} deformable mirror hardware from 6 \cite{douglas_wavefront_2018} to at least 7, demonstrating its operation in the \gls{LEO} space environment. 
DeMi will also demonstrate closed-loop wavefront control in space, correcting slowly-varying, high-order aberrations as well as quickly-varying tip-tilt errors. Closed-loop, in-space wavefront control with a MEMS DM is currently at a \gls{TRL} of 5 \cite{giveon_broadband_2007,rao_path_2008,bendek_development_2016}, and DeMi will also raise this to 7. The performance requirement for the DeMi payload is to measure the surface of the DM to an accuracy of 5 nm \cite{marinan_improving_2016}, requiring wavefront sensing accuracy of 10nm.

To complete the primary objective of demonstrating deformable mirror operation in space, the DM actuators must be commanded and their positions measured accurately. The DeMi payload includes a custom DM driver and a \gls{SHWFS} to measure the optical surface through wavefront reconstruction. 
There are two options for \gls{DM} illumination, an external aperture viewing a star, or an internal laser diode.

In this work, we present the design and development progress of the DeMi payload. In Section \ref{sec:optomech} we describe the requirements and approach for the optomechanical design. In Section \ref{sec:electrical} we provide an overview of the electronics system and details of its implementation, as well as an explanation of our wavefront sensing approach. In Section \ref{sec:lab_validation} we present the work done so far to validate the DM driver electronics design and wavefront sensing methods. Section \ref{sec:summary_and_future} contains a summary of work completed thus far and a discussion of remaining work needed to bring the project to completion. In this paper, we show maturation of the DeMi payload design, and significant progress towards integration and testing of the flight payload.

\subsection{DeMi Spacecraft Bus}
The DeMi CubeSat is based on the Blue Canyon Technologies XB6 platform, a 6U (30 cm x 20 cm x 10 cm) spacecraft bus comprising solar panels, batteries, command and data handling electronics, communication systems, and \gls{ADCS}.  For DeMi, the bus contains a Cadet U UHF radio capable of receiving commands from the ground and downlinking data to a 18 m dish at NASA Wallops Flight Facility at a rate of 1 Mbps. It will also have an Astrodev Lithium 2 UHF radio for backup communications at up to 9.6 kbps. The backup UHF ground station is located on the MIT campus in Cambridge, MA. The XB6 bus includes the XACT \gls{ADCS}, which is made up of a star tracker, an inertial measurement unit, magnetic torque rods, and reaction wheels. The system is capable of accurately pointing the bus with stability better than 10 arcseconds ($1\sigma$) in all axes \cite{mason_minxss_2017}. 

\section{Optomechanical Design}
\label{sec:optomech}
\subsection{Optical Layout} 
To characterize the operation of the 140 actuator \gls{DM}, a stable wavefront is required. 
To increase redundancy and allow cross-calibration, DeMi employs both image plane and pupil plane (Shack-Hartmann) wavefront sensing.  
Design of a relatively complex, many-element optical system for a CubeSat poses several challenges: a constrained payload volume (less than 4U of the 6U spacecraft\footnote{1U corresponds to a 10 cm x 10 cm x 10 cm cube}), a potential operating temperature range of -5$^\circ$C to 20$^\circ$C, and cost constraints.
To address  these challenges, a design based on \gls{cots} diamond-turned \gls{OAP}  mirrors is used. 
Off-the-shelf available focal lengths and apertures are chosen to enable rapid lab prototyping.
All \gls{OAP}s are catalog parts from Thorlabs Inc (Newton, New Jersey, USA). 
For the flight mirror assemblies, non-reflective surfaces are procured un-anodized and will be wrapped in polyimide tape for increased emissivity. 
Reflective surfaces will be aluminum with a SiO$_2$ protective coating.
Early designs\cite{marinan_improving_2016} employed singlet lenses which induced significant chromatic aberration when observing stellar targets in broadband.
Other designs including commercial, multi-element lenses\cite{douglas_design_2017-1}
corrected chromatic aberration while providing a large \gls{FOV},  but added significant mass and complexity, as well launch and thermal environment survival risk.

\begin{figure}[htbp]
\begin{center}
\includegraphics[width=0.75\textwidth]{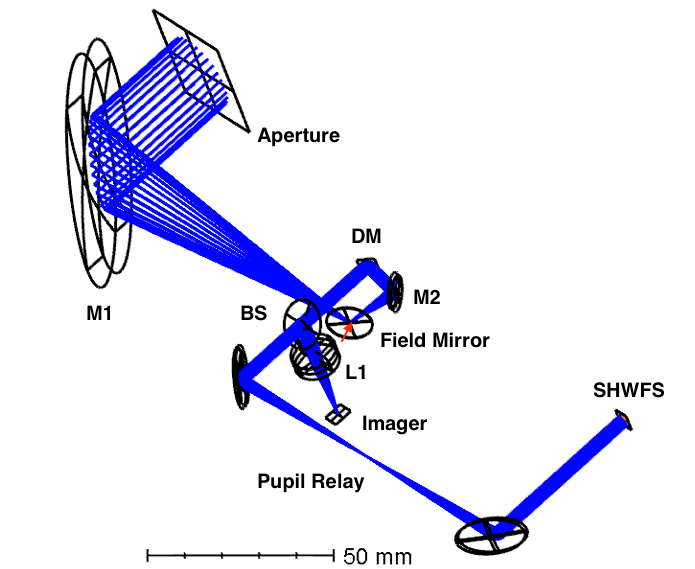}
\caption{Ray trace of the DeMi payload optical layout performed in Zemax for an on-axis stellar source, incident from the aperture in the upper right via a baffle (not shown). The red arrow indicates the site for injection of a laser source via a single-mode optical fiber in the center of the field mirror,}
\label{fig:optical_layout}
\end{center}
\end{figure}
The  layout shown in Figure \ref{fig:optical_layout} includes a telescope assembly, the deformable mirror, a pupil relay and two imaging cameras.
 The primary mirror is a 50.8 mm diameter, 100 mm focal length 90 degree \gls{OAP} , which will be under-illuminated by an input baffle, not shown.
A polished aluminum field mirror manufactured in two parts allows injection of monochromatic light into the image plane via single mode fiber.
A 12.7 mm diameter, 15 mm focal length OAP collimates the beam and illuminates the approximately 5 mm aperture of the deformable mirror in a pupil plane.
A pair of 50 mm focal length, 12.7 mm diameter OAPs relay this plane to the wavefront sensor. 
The wavefront sensor samples the beam with 150 um lenslets, providing better than Nyquist sampling of the \gls{DM} actuators. 
Before the relay, a non-polarizing  50/50 beamsplitter picks off light to the imaging camera, which is fed by a 25 mm focal length achromatic lens.
The imaging camera can function as either a passive monitor of wavefront state or an active wavefront sensor controlling the deformable mirror (see Section \ref{sec:wfs_test_results}).
This optical train provides diffraction limited image performance and  provides redundant sensing of the \gls{DM} surface.

\subsection{Optomechanical Tolerances}
The optical assembly is toleranced in Zemax using Monte Carlo simulations of alignment in each degree of freedom that impacts system wavefront error.
This provides the fine adjustment levels required to keep the misalignment contribution to the imaging system wavefront budget below  $\lambda/4$ to maximize the stroke available for active control. 

\subsection{Mechanical Design} 
The structural design of DeMi mainly focuses on adjustability of the position of optical elements so that they can be aligned. Repeatability of this alignment is also important, as is robustness of the assembly to mechanical vibration. The structure must have the ability to align the optical elements, to survive a 100g load in all three axes, and maintain alignment through as much of the expected -5$^\circ$C to 20$^\circ$C thermal range as possible; all while minimizing manufacturing costs. Built-in adjustment is necessary to aid in alignment during mechanical integration and tolerate variation in manufactured optics. 
Quick repeatability of alignment after disassembly is also required. The aluminum OAPs greatly simplify optic attachment when compared to glass optics whose coefficent of thermal expansion varies greatly from that of aluminum mounting brackets.
Aluminum optical bench and mounting structures are thermally connected to each other and the OAPs with thermally conductive gap filler.


For DeMi, kinematic mounting of optical components is achieved through the use of fine-pitch (Thorlabs M2.5 x 0.20) adjustment screws with ball ends, which are threaded into appropriate fine-pitch bushings. These are paired with socket-head machine screws with Belleville washers to provide retention force. The adjustment screws allow precise translation along their axis and, when combined in groups of 2 or 3, allow precise rotation of a mount or part in all three axes.

\begin{figure}
\centering
\includegraphics[width=0.70\textwidth]{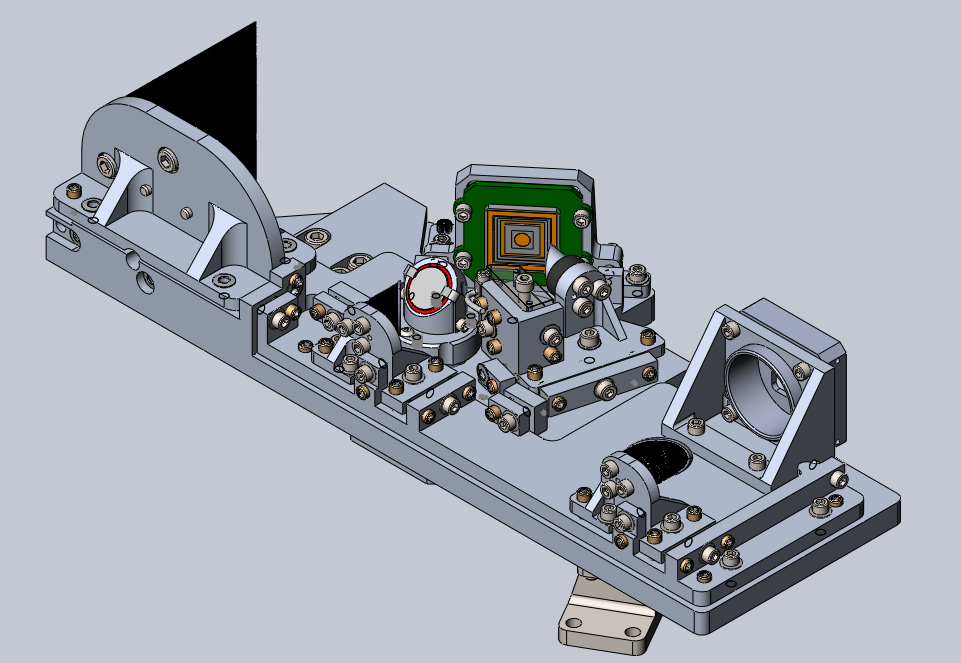}
\caption{\label{fig:SWassembly}CAD model of the proposed design.}
\end{figure}

To make the structure more rigid, the underlying optical bench is a monolithic piece. This base is attached to the bus through three 6Al-4V titanium stand-off flexures, which form an exact constraint and allow the payload to survive launch-induced vibrations. Five subassemblies are attached to the optical bench using kinematic mountings. These subassemblies hold in place one field mirror, four off-axis parabolic mirrors, the deformable mirror, one beam splitter, lenses, and cameras, as shown in Figure \ref{fig:SWassembly}. To make sure the focal lengths and position adjustments do not change with temperature variations, the mirrors and major structural components are all made of the same material, 7075 aluminum. In order to apply consistent torque to the fasteners, the fine-pitch screws are adjusted before the fasteners are torqued.

A finite element analysis was performed in SolidWorks to demonstrate that all interfaces would survive the vibration loads of the launch environment. The RMS vibration amplitudes were applied as a simulated 100g static load applied along each axis simultaneously. This load was chosen to be more conservative than the requirements of the NASA General Environmental Verification Specification (GEVS)\cite{milne_general_2003}, which represents a 3-sigma case of applied acceleration due to launch-induced vibrations.  The resulting reaction forces on each of the fasteners were analyzed for failure modes such as bolt shear and tension failure, bolt head pull-through, and bolts near part edges shearing through the sides of their holes. The preload tension required to to prevent gapping between parts under these loads requires fasteners made of stainless steel with more than 100 ksi tensile strength, such as A286. Results of this analysis are shown in Table \ref{tab:fastener_results}. All failure modes except gapping have margins greater than a factor of 3. As a non-critical failure, a safety factor above 1.8 was deemed to be acceptable for gapping.

\begin{table}[]
\centering
\begin{tabularx}{0.8\textwidth}{ |c| *{7}{Y|} }
\hline
Failure Mode & Bolt Shear & Bolt Tension & Gapping & Head Pull-through     & Plate Shear-out & Net Tension      \\
                \hline
Margin    & 5.76       & 3.96         & 1.875   & \textgreater{}10 & 3.24            & \textgreater{}10 \\
\hline
\end{tabularx}
\smallskip
\caption{\label{tab:fastener_results} Results of fastener analysis based on a 100g load on each part in each axis. Each value reflects margin above failure stress for the worst-case fastener in the assembly for each particular failure mode. Margin has been calculated such that a value of 1.0 indicates the part is at failure.}
\end{table}
\subsubsection{3D Printed Model}

\begin{figure}
\centering
\includegraphics[width=0.70\textwidth]{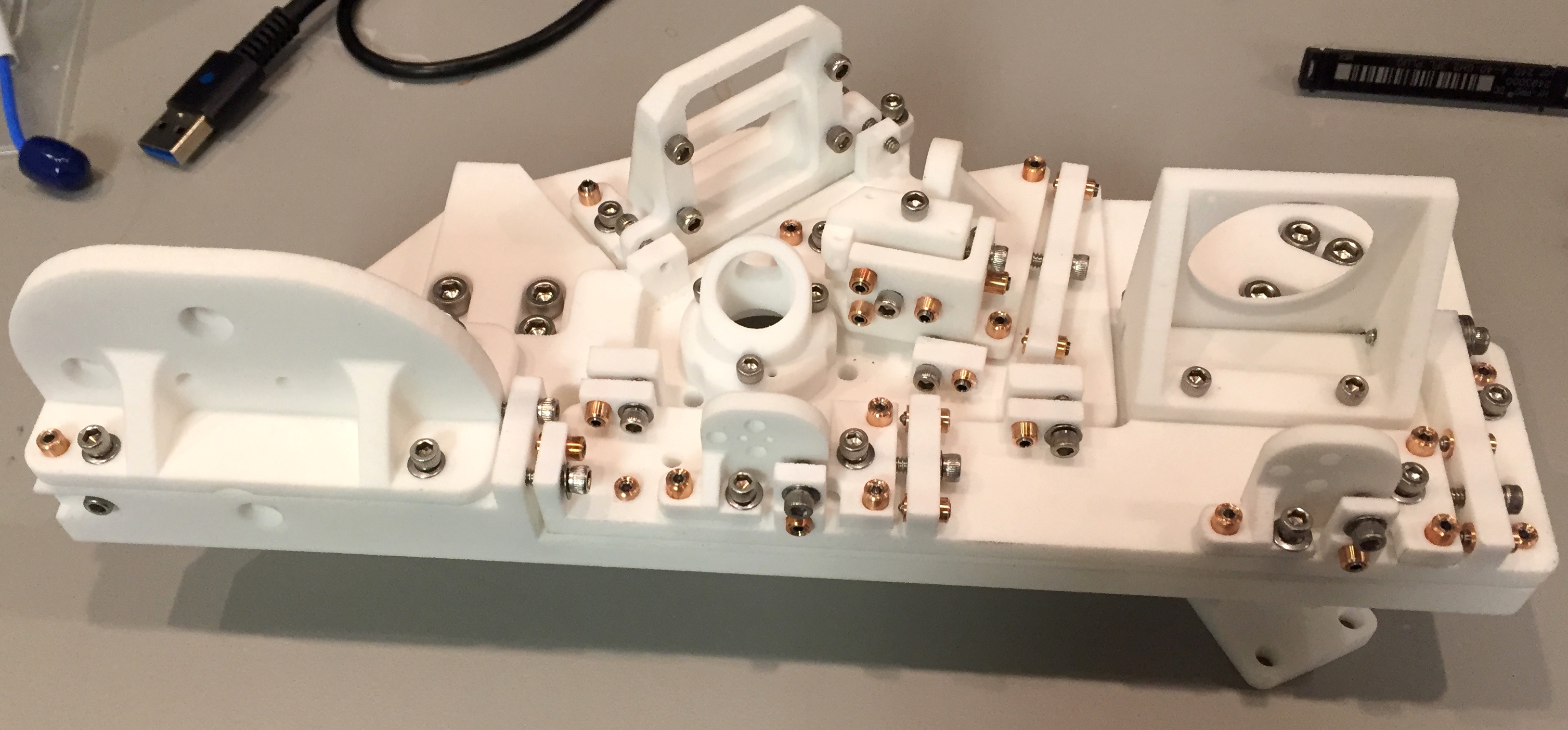}
\caption{\label{fig:3DModel1}Assembled 3D print of the payload without mounted optics. Photo taken in March, 2018.}
\end{figure}

\begin{figure}
\centering
\includegraphics[width=0.70\textwidth]{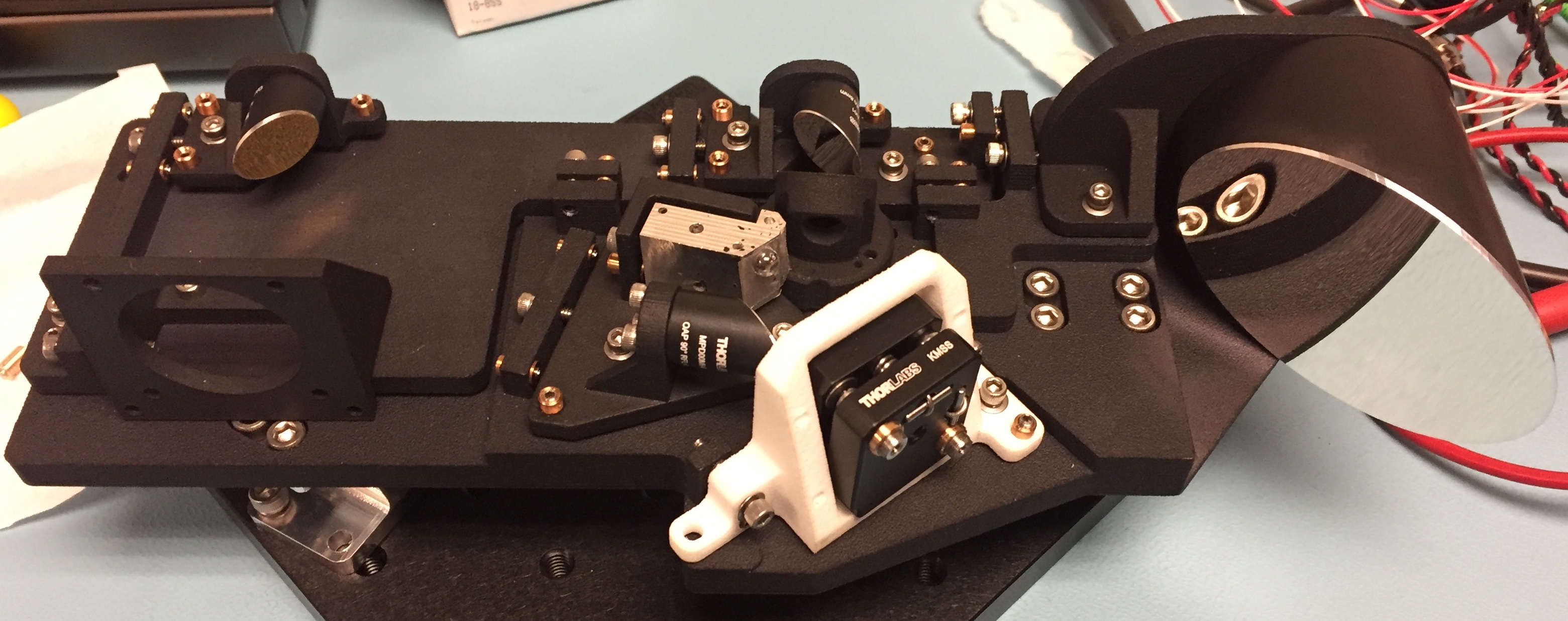}
\caption{\label{fig:3DModel2}The second iteration of the 3D printed model with the optics mounted. Photo taken in May, 2018.}
\end{figure}

A prototype of the DeMi structure was 3D-printed in order to perform fit-checks and test the optical alignment procedure prior to final fabrication. Some adjustments were made after initial fit-checks so a second version was 3D-printed. Optical elements were installed in the 3D-printed model and initial attempts at coarse alignment of the optics were made using a Zygo VeriFire QPZ interferometer. Alignment was not performed to a satisfactory level, as the printed model is not tightly toleranced and the adjustment screws do not fit tightly in all the parts. However, this served as a valuable rehearsal of alignment procedure, and based on these experiences, minor changes to the alignment adjustment mechanisms have been incorporated into a final design. Specifically, steel pins were added to act as contact points for the ball-tipped adjustment screws to prevent deformation of the aluminum surfaces. At this time, the design has been sent to the Boston University Scientific Instrument Facility for fabrication of the flight version in aluminum. Figure \ref{fig:3DModel1} shows the first version of the prototype and Figure \ref{fig:3DModel2} shows the second prototype with representative COTS OAPs mounted.

\subsection{Thermal Analysis}


\begin{table}[h]
\centering
\begin{tabularx}{0.8\textwidth}{ |c| *{5}{Y|} }
\cline{2-5}
\multicolumn{1}{c|}{} & \multicolumn{2}{c|}{Survival Temp. Range ($^\circ$C)} & \multicolumn{2}{c|}{Operational Temp. Range ($^\circ$C)} \\
                   \hline
Component          & Min                   & Max                  & Min                    & Max                    \\
                   \hline
Deformable Mirror  & 0                     & 80                   & -10                    & 30                     \\
Raspberry Pi CM3s  & -25                   & 85                   & -25                    & 85                     \\
DM Driver Boards   & -40                   & 50                   & -40                    & 50                     \\
Power Supply Board & -55                   & 65                   & -55                    & 85                     \\
PixeLink Cameras   & -45                   & 85                   & 0                      & 50                     \\
Optics Bench       & 5                     & 35                   & 16                     & 24                     \\

\hline
\end{tabularx}
\smallskip
\caption{\label{tab:thermal_limits} Survival and operational temperature limits for DeMi payload components derived from component datasheets. The aluminum optical bench structure has the tightest requirements for both regimes.}
\smallskip
\end{table}

Thermal analysis was performed for the DeMi payload to assess the temperature ranges that components will experience during all stages of the mission. In order to maintain negligible ($<10 $ nm wavefront error) thermoelastically induced misalignments on optical elements, all optical elements and the bench are required to maintain their absolute temperature to $20^{\circ}\text{C}\pm4^{\circ}\text{C}$ during operational periods (up to 1000 seconds) and temperature gradients must be less than 2$^\circ$C between optical components.
 The DeMi payload must maintain temperatures within the survival ranges of all components at all times. Temperature limits for each component are shown in Table \ref{tab:thermal_limits}. The  payload uses thermal interface material to create known thermal interfaces between kinematically mounted components, and uses two  2.5 W heaters (located on the underside of the optical bench) to actively maintain component temperature requirements.

A thermal model was developed by Blue Canyon Technologies to model predicted spacecraft wall temperatures for hot and cold extremes of orbital conditions, as determined by solar beta angle and exact orbital parameters. Using this data, five cases are identified as bounding thermal cases and are modeled in Thermal Desktop: hot operational, hot storage, cold operational, cold storage, and commissioning. In the operational cases the heaters are used if necessary and the payload electronics are active and dissipating heat. In the storage cases the heaters are used if necessary, but the electronics are powered off. The commissioning case captures conditions immediately after spacecraft deployment before heaters or electronics are powered on.  
These five cases were simulated in Thermal Desktop using the bus wall temperature data as boundary conditions, modeling conduction paths within the payload and to the bus, internal radiation (both payload to host and payload to payload), component power dissipation, and heater control laws. 

\begin{figure}[h]
\centering
\includegraphics[width=0.80\textwidth]{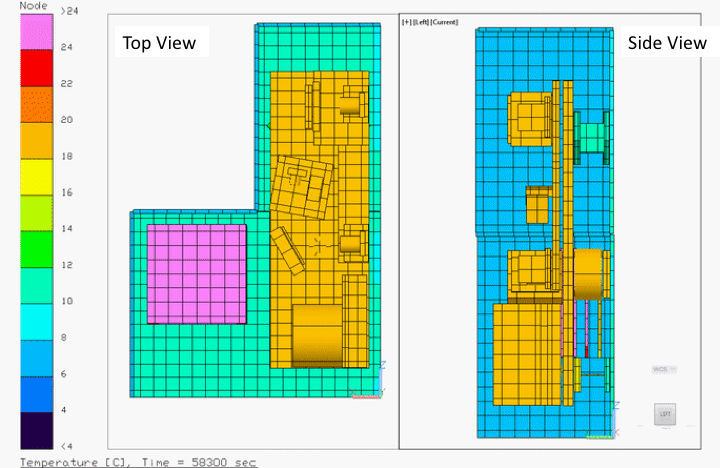}
\smallskip
\caption{\label{fig:demithermalmodel}Top and side view of DeMi Thermal Desktop model results at a single time step during a hot case simulation near the end of an observation during eclipse. Electronics boards (left), payload optical assembly, and spacecraft walls can be seen. In this simulation, the optical bench maintains a uniform temperature to within one degree C. }
\end{figure}

The input wall temperatures and the results of this analysis are shown in Table \ref{tab:thermal_results}. For all cases examined, the DeMi payload components remained within their required temperature limits. Optomechanical temperature requirements refer to those of the bench and DM, while the electronics requirements refer to the other components in Table \ref{tab:thermal_limits}. In the commissioning case, the optical bench comes within 1$^\circ$C of its predicted survival range without permanent misalignment of optical components. More analysis and testing are required to verify the thermal model and the bench's survival temperature. Thermal stability during operational periods was found to be $\pm 1^{\circ}$C across the optical bench and mounted optical components, and all components remained within their survival temperature range both in storage and commissioning phases of the mission. Figure \ref{fig:demithermalmodel} displays a temperature map of the DeMi payload for a single time step in the hot operational case.

\begin{table}[h]
\centering
\begin{tabularx}{0.8\textwidth}{ |c| *{6}{Y|} }
\cline{2-6}
\multicolumn{1}{c|}{} & Conditions ($^\circ$C) & \multicolumn{2}{c|}{Results ($^\circ$C)}  & \multicolumn{2}{c|}{Limits ($^\circ$C)}                 \\
\hline
Test Case            & SC Wall Temp. & Optomech. Temp. & Electronics Temp. & Optomech. Temp. & Electronics Temp. \\
\hline
Hot Storage          & 3 - 18               & 10 - 12            & 9 - 13        & 5 - 35  & -25 - 50        \\
Hot Operational      & 3 - 18               & 18 - 20            & 10 - 48       & 18 - 22 & -25 - 50           \\
Cold Storage         & -3 - 13              & 10 - 15            & 3 - 9         & 5 - 35  & -25 - 50           \\
Cold Operational     & -3 - 13              & 18 - 20            & 7 - 45        & 18 - 22 & -25 - 50           \\
(Cold) Commissioning & -3 - 13              & 5 - 6              & 7 - 45        & 5 - 35  & -25 - 50           \\               
\hline
\end{tabularx}
\smallskip
\caption{\label{tab:thermal_results} Results of the Thermal Desktop Simulation. All components are maintained within their thermal limits. In the commissioning case where heaters are not active, the optical bench is predicted to be at its survival limit.}
\smallskip
\end{table}

\subsection{Environmental Testing}
Preliminary thermal cycling tests have been performed on DeMi off-axis parabolic mirrors, which were not rated by the manufacturer for the expected on-orbit temperature range. These Thorlabs aluminum mirrors have a SiO\textsubscript{2} optical coating and black anodization on non-optical surfaces. In order to determine if these coatings would remain adhered to the mirrors after thermal cycling, a test unit was repeatedly cycled between room temperature and 60$^{\circ}$C. Adhesive tape was then applied and removed in an attempt to induce the coatings to flake, and both coatings survived this test without damage. However, the outgassing properties of the dyes used in the black anodization are unknown, so the flight parts will be non-anodized but will be wrapped in polyimide tape to maintain thermal emissivity in an acceptable range.

Full functional testing will be performed on the integrated DeMi payload in a vacuum environment over the expected spacecraft wall temperatures of -3$^\circ$C to 18$^\circ$C to replicate the expected hot and cold cases in Table \ref{tab:thermal_results}. Vibration tests will also be performed on a vibration table with a fixture to replicate attachment to the spacecraft bus, at low levels of ~4g RMS in order to verify workmanship. Full functional testing will be performed before and after vibration. Once the payload is integrated into the spacecraft, it will be subjected to environmental testing including Thermal Vacuum (TVAC), shock, and vibration in accordance with the NASA GEVS \cite{milne_general_2003} or applicable launch service provider requirements. The spacecraft will be treated as a protoflight article and so will be tested to qualification levels for a reduced duration.

\section{Electrical Design}
\label{sec:electrical}
\begin{figure}
\centering
\includegraphics[width=0.80\textwidth]{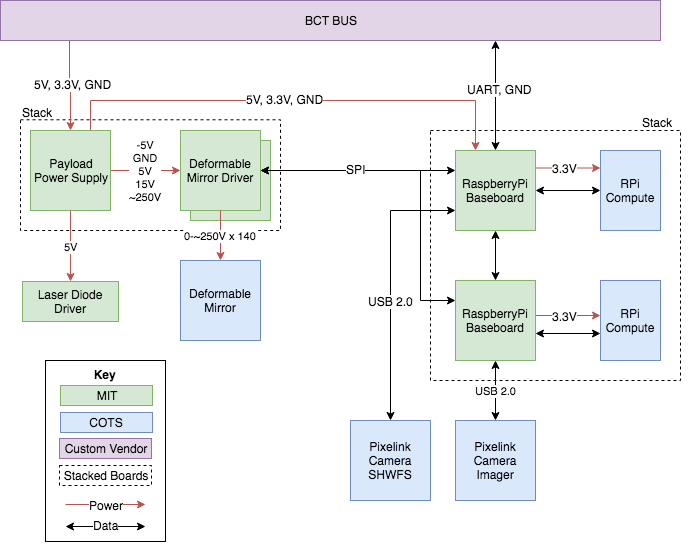}
\smallskip
\caption{\label{fig:demi_electronics}Block diagram of the DeMi electronics system. Most boards are part of one electronics stack, as indicated by the dotted line. The stack is held together by standoffs between boards, with header connectors to provide electrical connections along the stack. The key shows which components are designed by the DeMi payload team at MIT and which come from external vendors.}
\end{figure}

The DeMi payload electronics stack consists of a power distribution board, two DM driver boards, and two processor boards. These boards are configured as a stack in a sub-1U form factor with stacking board-to-board electrical connectors precluding the necessity for major inter-component harnessing. Ribbon cables will be used to connect the DM driver boards to the DM. Power will be provided by the XB6 bus in three channels that can be independently switched by bus commands; one at 5V and two at 3.3V. High-voltage (250V) and -5V power for the DM controller will be generated on the power distribution board. A constant-current power supply will be mounted beside the stack and used to drive the internal laser diode. A block diagram of the DeMi payload electronics is shown in Figure \ref{fig:demi_electronics}.

\subsection{Payload Computer} 
The payload computer consists of two Raspberry Pi Compute Module 3  computers hosted on custom carrier boards. These computers run the Raspian Linux OS, which includes facilities for direct control of the Raspberry Pi input/output pins. The payload computing system is split between these two boards, which have redundant functionality for fault tolerance purposes. One board primarily handles readout from the science imager while the other handles the Shack-Hartmann wavefront sensor data. As described in Section \ref{ssec:wf_sns_control}, the image plane sensor and the SHWFS are both capable of monitoring the performance of the DM, while each also performing secondary tasks.  Both computers are also capable of interfacing with the DM driver electronics via a Serial Peripheral Interface (SPI) connection, with the power supply board via UART, and each can communicate independently with the spacecraft bus for commanding and telemetry purposes. These capabilities mean that if one camera or Compute Module completely fails, the DM can still be operated and monitored, fulfilling DeMi's primary objective.

Each Compute Module carrier board also has built-in redundant program memory. In case of a fault on one Compute Module, a hardware watchdog timer present on the carrier board is capable of resetting the Compute Module and configuring the hardware to boot from a separate on-board SD card. This means that a boot image corrupted by a single-event upset in FLASH memory will not result in the Compute Module becoming permanently inoperable.  When the watchdog timer is triggered it will latch a thyristor and permanently engage the secondary memory card via a multiplexer (Fairchild FSSD06). The watchdog will also trigger a momentary deactivation of the power supply to the compute board to initiate a reset.  This reset circuitry is powered directly from the bus 3.3V lines, which are independently switched on the bus. This full and independent power control from the bus also gives the ability to reset the latched Thyristor -- if reversion to the original SD card is desired.  

\subsection{Deformable Mirror Controller}
Due to the large size of \gls{COTS} \gls{DM} controllers, a new custom driver was designed for DeMi. 
The DeMi deformable mirror controller consists of two separate high voltage driver boards, each consisting of three driver units. Each unit implements 32 driver channels controlled by a single SPI bus interface and is composed of a 32 channel 14-bit Digital-to-Analog Converter (DAC) chip (Analog Devices AD5382), 32 channel high voltage amplifier (Microchip-Supertex HV256), a high side current monitor chip (Microchip-Supertex HV7802), a digital temperature readout chip (Maxim MAX6627) and a general-use Analog-to-Digital Converter (ADC) chip (Analog Devices AD7680). The 140 actuators could be accommodated with just 5 of the 32-channel control units, but for ease of fabrication and testing it is preferable to assemble two identical 3-unit boards.

\begin{figure}
\centering
\includegraphics[width=0.70\textwidth]{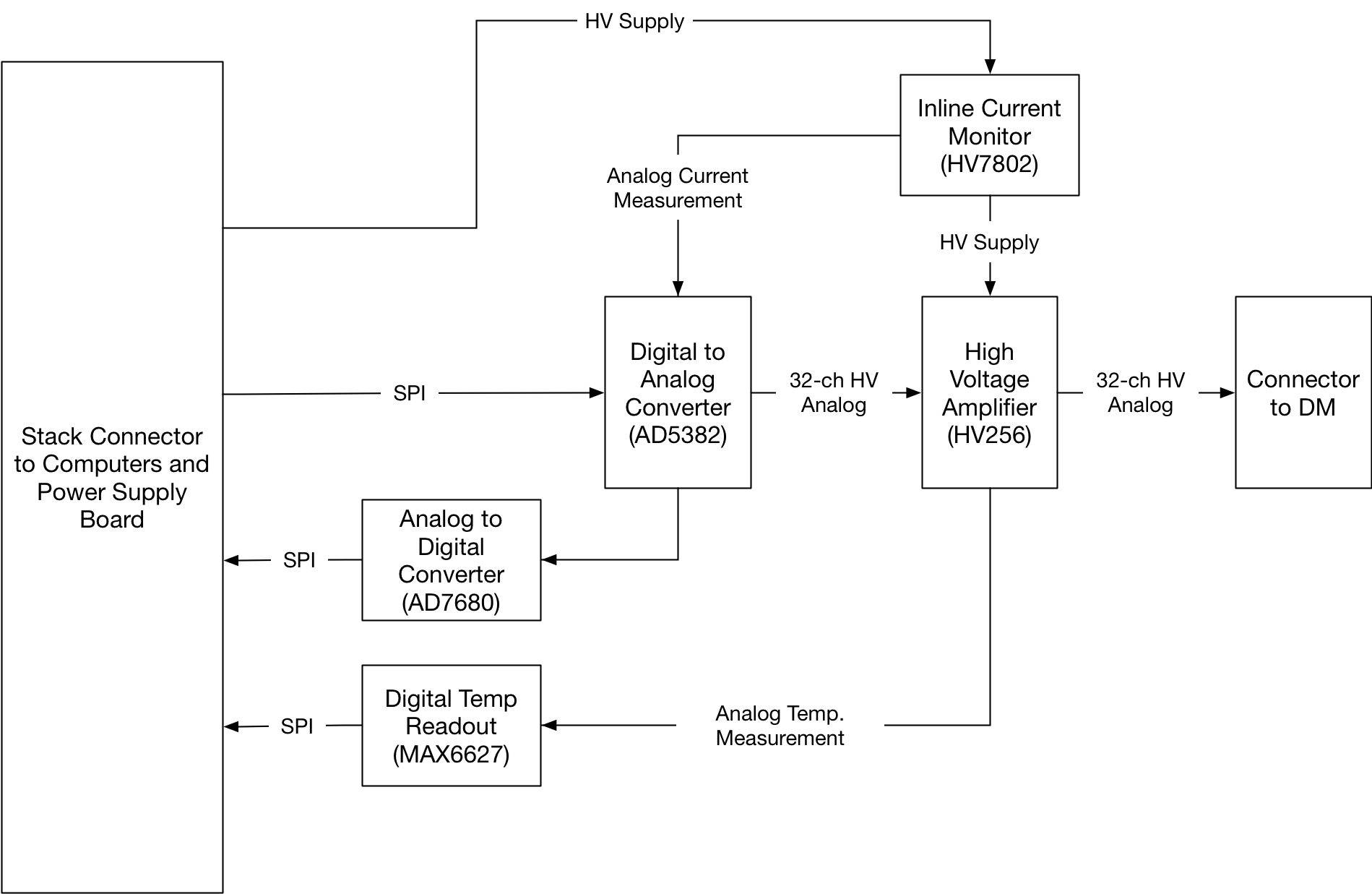}
\caption{\label{fig:DM controller}Block diagram of a 32-channel high-voltage DM controller unit. Each custom controller board contains 3 of these units and the DeMi electronics stack has two of these boards to accommodate the 140-channel DM.}
\end{figure}

To control the DM, the microprocessor writes values to the DAC through an SPI interface. The low voltage (LV) analog signals from the DAC are amplified by the high voltage (HV) amplifier, the output of which is then routed to the DM. Current draw from the HV supply for each unit is measured by an inline current monitor which produces an analog value proportional to current consumption. The DAC chip also includes a 36 channel multiplexer which can route the analog voltage from any of the 32 LV outputs or any of 4 analog input pins to a single buffered output pin. Therefore, the current measurement signal is routed to one of the DAC's 4 analog input pins and a generic SPI-enabled ADC is connected to the output pin. The HV amplifier IC has a built in silicon temperature sensor on its die which is read by a dedicated readout IC which is also interfaced to the microprocessor via SPI. A block diagram showing the components of a single unit is included in Figure \ref{fig:DM controller}.

A 32-channel DM controller unit has been prototyped in hardware and demonstrated with a Raspberry Pi and smaller DM as discussed in section \ref{ssec:elec_test}.

\subsection{Wavefront Sensing and Control Approach} 
\label{ssec:wf_sns_control}
The DeMi payload will be capable of measuring full wavefront shape with both an image-plane sensor and a pupil-plane \gls{SHWFS}. Both sensors are based on the same \gls{COTS} industrial CMOS camera, the Pixelink PL-D775MU-BL, with an ON Semiconductor MT9P031 5-megapixel sensor. The SHWFS uses a Thorlabs MLA150-5C microlens array to create the necessary spot field. Each sensor is also capable of measuring tip-tilt errors in the wavefront at a higher rate to enable separate closed-loop control of high-frequency disturbances. Wavefront control is performed using a hybrid approach, where separate control loops are used for high-rate tip-tilt correction and low-rate high-order correction. Control data from both loops are then combined to drive the DM surface. 

Image plane wavefront sensing will be performed using the differential optical transfer function (dOTF) approach \cite{codona_differential_2013}. This requires that a perturbation be added to the wavefront near the edge of the aperture, which can be easily accomplished using an actuator of the DM. Tip-tilt errors can be quickly determined in the image plane using displacement of the PSF in a central region of image pixels. Shack-Hartmann wavefront reconstruction will be performed using a zonal fitting method by which a phase surface with the same resolution as the wavefront sensor will be fit to the sensor's local gradient data \cite{southwell_wavefront_1980,herrmann_least-squares_1980}. Tip-tilt errors can be measured by the SHWFS by averaging displacements over a single row and column of lenslet spots. Full wavefront sensing with both methods has been demonstrated on lab testbeds as discussed in section \ref{sec:wfs_test_results}. With both sensors, the full wavefront reconstruction methods are very sensitive to the values of physical parameters of the optical system. Errors in these values will be corrected by comparing the results with known wavefront displacements. The  known displacements will be provided by actuating the flight DM to the shape of various zernike polynomials using a controller which has been calibrated by the vendor for this purpose.

\begin{figure}[t]
\centering
\subfloat[]{\includegraphics[width=0.67\textwidth]{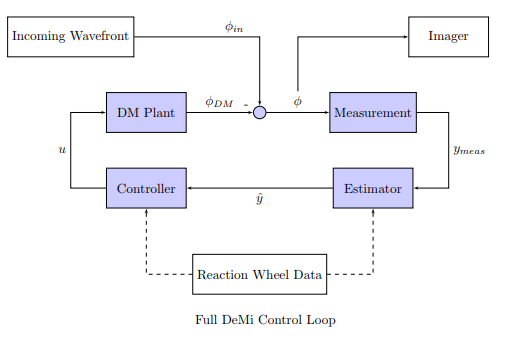}}

\subfloat[]{\includegraphics[width=0.8\textwidth]{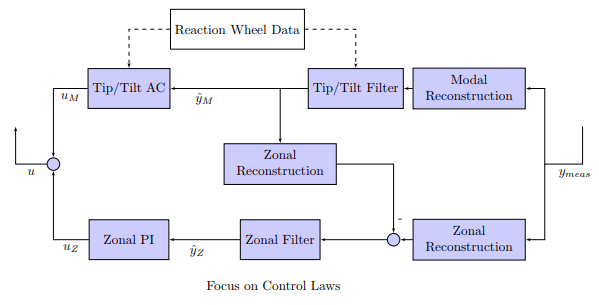}}
\smallskip
\caption{\label{fig:Dual_control_loop} Simplified block diagram of full DeMi control loop (a), showing DM correction of wavefront errors from wavefront measurements with input from reaction wheel data. The controller is shown in more detail in (b) incorporating  the dual estimation and control loop to control high temporal frequency tip/tilt at a higher rate than the higher spatial frequency disturbances. The zonal control blocks are modeled with MATLAB scripts to verify the the rejection of high spatial frequency disturbances.}
\end{figure}

The capability to measure the wavefront with both sensors sensors means that both high-rate tip-tilt correction and the primary mission of DM validation can still be performed in the case of failure of a single payload processor board. In the absence of any failures, primary wavefront control will be accomplished on orbit using data from the SHWFS. Tip-tilt control is planned to run at a rate of 36 Hz, which is primarily limited by readout of sensor data over USB 2.0 and subsequent processing into wavefront error signals. Taking advantage of the single-dimensionality of first order disturbances, DeMi processing only requires one full row and column of SHWFS data to estimate the tip tilt disturbance.  This allows the loop to run at a higher rate despite being limited by the computer and sensor's data readout bandwidth.  Full high-order control will be performed in a separate thread at a lower rate, 1-3 Hz, with the full wavefront zonal reconstruction results.  An estimate of the tip-tilt correction will be taken out of the full zonal reconstruction to avoid negative interactions between the different control laws running at 36 Hz and 1-3 Hz.  The full zonal control for the higher spatial frequency modes is done using a proportional integral (PI) control law, while the tip tilt controller uses an adaptive control (AC) law that adapts to the current speed of the reaction wheels.  This approach is shown graphically in Figure \ref{fig:Dual_control_loop}. In the case of a failure in the SHWFS or its payload computer, the tip-tilt control can be based on PSF displacement in the image plane. However, more work is necessary to determine if image-plane wavefront sensing can be performed quickly enough for use in high-order closed-loop wavefront control.

The hybrid control approach was chosen because the spacecraft's reaction wheels will cause tip-tilt disturbances which are low spatial frequency, but high temporal frequency (main modes up to 12 Hz). Disturbance modeling was performed as described in Section \ref{sec:pointing_modeling}, with the assumption that the reaction wheels used on DeMi are similar to the BCT 15 wheels analyzed for the ASTERIA mission\cite{shields_characterization_2017}.  By sampling only a subset of the sensor pixels for tip-tilt sensing, the data bottleneck can be overcome to allow the control law to run at more than double the expected speed of the main disturbance source.  Any higher spatial frequency disturbances will have a low temporal frequency (periods on the order of 5 to 90 minutes) because they are driven mostly by the thermal effects from DeMi mode transitions and/or orbital location (e.g., going from eclipse to sunlight).

\subsection{Pointing Modeling}
\label{sec:pointing_modeling}
\subsubsection{Modeling Overview}
In order to determine whether the two-stage hybrid control approach is necessary DeMi, the DM plant, disturbances, estimator, and control law are simulated using MATLAB and Simulink.  The DM plant is modeled with each actuator as a 2D Gaussian that affects the nearby actuators to account for a physical coupling factor.  The response of each actuator is based on representative test curves for the 5.5 micron Multi-DM provided by Boston Micromachines. The tip-tilt disturbances are modeled with the static imbalance from the reaction wheels as the primary source of tip-tilt.  This model is generated from BCT 15 attitude control system test data, specifically the amplitude and disturbance frequency as a function of wheel speed in RPM \cite{shields_characterization_2017}. For the higher order spatial frequency terms, several different disturbance waveforms are used that directly affect the wavefront measured by the simulated SHWFS.  These waveforms are generated by using time-varying Zernike expansion coefficients (up to fourteenth order), with individual peak to valley amplitudes as high as 3.5 microns.  Five different Zernikes at a time are chosen, their amplitudes modulated by waveforms with periods varying from five to ninety minutes, and combined to form a single disturbance waveform.
\subsubsection{Slow High-order Disturbances}
\begin{figure}[!h]
\centering
\includegraphics[width=0.6\textwidth]{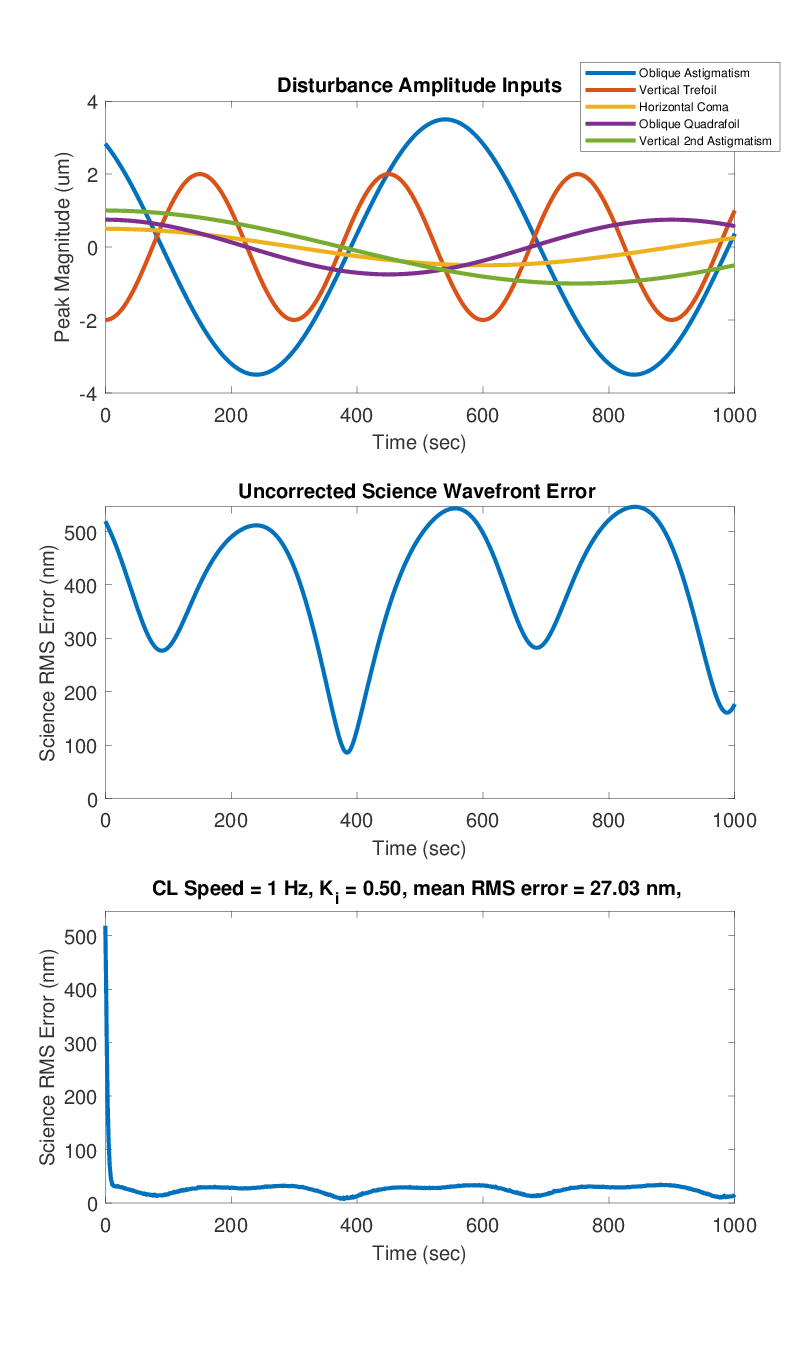}
\caption{\label{fig:zonalControl_higherOrder} Results of simulating a 1 Hz zonal control loop against high spatial frequency disturbances shown in the top plot. Middle plot shows the varying RMS error at the image plane if no control is used. The bottom plot shows science wavefront error correction, with control integrator gain, and error mean.  The resulting mean RMS wavefront error is 27.03nm, with RMS error below 25 nm over 95\% of the 1000 second simulation time. }
\end{figure}

The higher spatial frequency disturbances are first simulated on their own, corrected only with a zonal control law at 1Hz.  The results from a 1000 second simulation are shown in Figure \ref{fig:zonalControl_higherOrder}.  The top plot shows the amplitude of each of the five disturbances, the middle plot shows the uncorrected RMS wavefront error at the science image plane and the bottom plot shows the corrected error. As shown in the bottom plot, after an initial startup error, the controller is fast enough to track the disturbances effectively and reject them with a mean RMS error of approximately 27 nm. This satisfies the DeMi payload requirement of wavefront correction to better than 100 nm RMS error.
 
\subsubsection{Fast Tip-tilt Disturbances}
When tip-tilt disturbances are added, the fastest rate the zonal control can be utilized to correct for them is estimated at 20 Hz. This could be accomplished by sampling 2 x 2 regions of pixels instead of each pixel. Running the zonal control at 20 Hz with the expected reaction wheel disturbance at 200 RPM and 1000 RPM leads to a mean RMS error over the simulation of 44 nm (for 200 RPM) and 47 nm (for 1000 RPM). The magnitude of disturbances is higher for the lower RPM value, which is why the mean error values are similar even though the lower RPM value has higher attenuation from the controller.  Note that PI gains have been optimized by sweeping through several gain values to determine lowest mean RMS value over the simulation.  These results lead us to the adaptive model control approach for tip-tilt to better address those disturbances by: 1) speeding up the loop by reading out only two rows and columns to estimate tip tilt,  and 2) using information from the current reaction wheel speed to better match the frequency and amplitude of the disturbance.

\section{Laboratory Validation}
\label{sec:lab_validation}
\subsection{DM Driver Test Results}
\label{ssec:elec_test}
Design validation is performed on a 32-channel DM controller prototype. The testing configuration utilizes an external High Voltage power supply, a Raspberry Pi 3B (Pi) and a 32 actuator bench-top deformable mirror (normally used with the large form factor vendor-provided driver) to validate the functionality and baseline the performance of the new driver.  The functionality of the driver and the accompanying sensing devices are confirmed through these tests.  The output levels of the driver are directly observed with a Fluke multimeter at a range of values to confirm the expected high voltage values are being delivered (0-250V). With a calibrating laser aligned at the center of the CMOS sensor, the DM is commanded to induce a linear tip and tilt on the wavefront, and the resulting displacements are observed in the image to be in the expected direction.  Each actuator is then individually driven to 50\% (with all other actuators relaxed at 0V such that the inter-actuator effects are minimal) and the resulting distortions are qualitatively observed to confirm functionality of each individual driving line and actuator channel. Following wavefront sensor calibration, additional tests will be performed to quantitatively measure the resulting wavefront motion from the driver's signals and to calibrate the driver accordingly. 

Prior to integrating the prototype driver, a proof of concept closed-loop controller is demonstrated to show the capability of the available benchtop deformable mirror for rejecting oscillation and introduced disturbances.  For the purposes of this demonstration, a tip-tilt disturbance on the order of half the experimentally determined throw range of the deformable mirror was introduced at 1 to 4 Hz. The Raspberry Pi 3 is used to compute the x and y centroid of the laser on the CMOS sensor at each frame at a 4 Hz rate. The centroid position is then used as error feedback to a proportional-integral control law in Matlab which interfaced with the vendor's DM driver.  This demonstration is promising for the implementation of the same demonstration with the prototype payload computer which we can control at a much faster and more consistent rate than in the demonstration configuration. The initial demonstration is limited to 4 Hz by the speed of the Matlab loop and communication overhead between the Pi and Matlab computer.

\subsection{Wavefront Sensing Test Results}\label{sec:wfs_test_results}
A representative optical system containing a \gls{DM}, image-plane sensor, and \gls{SHWFS} is used to perform validation of wavefront sensing methods. The objective of this test was to verify that a single DM actuator poke could be qualitatively seen in the reconstructed wavefront for each wavefront sensing method.
The light source for this test was a single mode fiber laser at  the image plane of a Vixen A80Mf refracting telescope used as a collimator and beam expander.
 The flatness of the resulting wavefront is verified manually with a shear plate. The beam is then reflected by a Boston Micromachines Mini DM before being split and directed to a flight-like image sensor and \gls{SHWFS}. 
To provide initial spot centroid positions for the SHWFS, data was initially collected with the DM in a relaxed state. Then an actuator near the center of the DM is chosen and actuated with 75V, inducing a single poke on the DM surface. Data is collected on both the SHWFS and image plane sensor in this configuration. Next, an additional actuator is poked near the edge of the DM so additional image-plane data could be collected to enable dOTF processing.

\begin{figure}[h]
\centering
\subfloat[]{\includegraphics[width=0.35\textwidth]{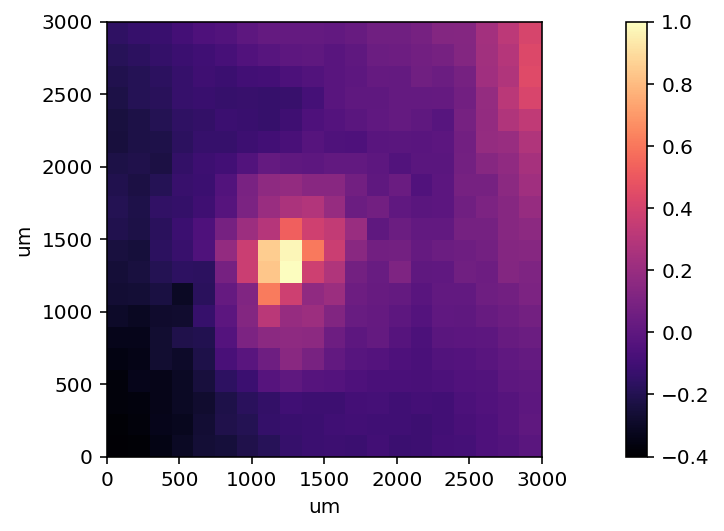}}
\subfloat[]{\includegraphics[width=0.35\textwidth]{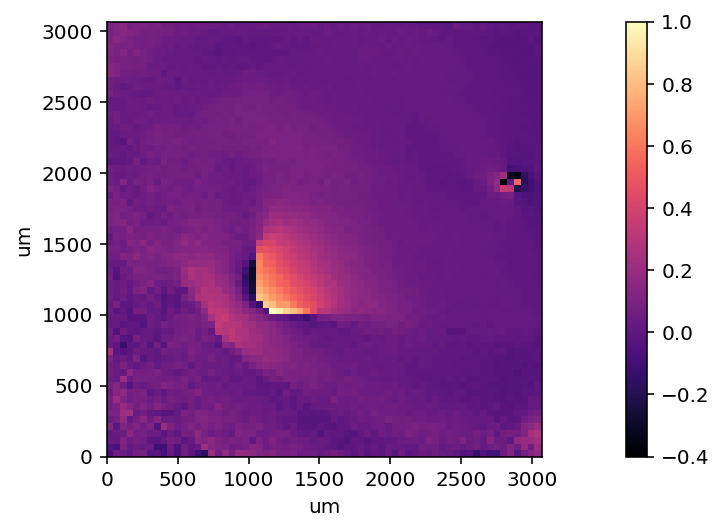}}
\caption{\label{fig:WFSTestResults} (a) Reconstructed wavefronts from laboratory tests using \gls{SHWFS} data with a zonal least-squares method and (b) image-plane data with the dOTF method. Both show a distinct actuator poke as expected. Color scales show wavefront displacement normalized to the maximum amplitude of the poke in each case. }
\end{figure}

Results are shown in Figure \ref{fig:WFSTestResults}. Both methods show a distinct actuator poke in the expected location. Note that the dOTF results are cropped to show a single version of reconstructed wavefront. Color scale is shown on a relative scale; additional work is required to calibrate both methods to the true amplitude of the wavefront as described in Section \ref{ssec:wf_sns_control}.

\section{Summary and Future Work}
\label{sec:summary_and_future}
The DeMi payload design demonstrates several key components of wavefront sensing and control on a small satellite platform.
The design of the mechanical and optical hardware has evolved toward the procurement of the flight unit and integration and testing. Mechanical fastener analysis shows design margins of at least a factor of 3 for critical failure modes. Thermal simulations show all components are kept within survival limits for the duration of the mission, and within operating limits when heaters are active.
The design of the electronics system is presented, and the higher-risk design of the DM controller board has been prototyped and passed basic functional checks, though more work is necessary to measure performance with a calibrated sensor. 
Our approach to adding redundancy in the payload processing and control system is presented. Though the sensors on each payload computer are different, their functionality overlaps to enable fault tolerance on orbit.
Using a laboratory optical testbed and numerical simulations, the wavefront sensing and control approaches are demonstrated; calibration with a known reference wavefront is planned.

The flight mechanical hardware design has been sent for manufacturing; assembly and alignment of all flight elements is planned. The electronics boards are in the fabrication process, with assembly, board-level and payload-level testing planned. Wavefront sensing techniques have been validated and are under refinement, and flight firmware and software implementations are in progress. Procedures for integrated payload testing and refinement of algorithms and operations are under review prior to delivery for launch in 2019.

\acknowledgments 
We acknowledge useful discussions with Jared Males of the University of Arizona, Jason Stewart of MIT Lincoln Laboratory, and Paul Bierden and Michael Feinberg of Boston Micromachines. 
 
A.J.S. participated as part of the Massachusetts Institute of Technology Undergraduate Research Opportunities Program.

This work at MIT has been sponsored by DARPA  under a contract with Aurora Flight Sciences, a Boeing Company.


\bibliographystyle{spiebib} 

\bibliography{DeMi,report} 

\end{document}